\documentclass{jetpl}
\twocolumn
\title{Random anisotropy 
disorder in superfluid $^3$He-A in aerogel}

\rtitle{Random anisotropy 
disorder in superfluid $^3$He-A in aerogel}

\sodtitle{Random anisotropy 
disorder in superfluid $^3$He-A in aerogel}

\author{
G.\,E.\,Volovik\/\thanks{e-mail: volovik@boojum.hut.fi}}

\rauthor{G.\,E.\,Volovik}

\sodauthor{Volovik}

\address{ Low Temperature Laboratory, Helsinki University of
Technology,
P.O.Box 2200, FIN-02015, HUT, Finland\\~\\
Landau Institute for Theoretical Physics RAS, Kosygina 2,
119334 Moscow, Russia}

\dates{18 September 2006}{*}

\abstract{The anisotropic superfluid $^3$He-A in aerogel provides an
interesting example of a system  with continuous symmetry in the presence
of random anisotropy disorder. Recent NMR experiments allow us to discuss
two regimes of the orientational disorder, which have different 
NMR properties. One of them, the (s)-state,
is identified as the pure Larkin-Imry-Ma state. The structure of another state, the (f)-state, is not
very clear: probably it is the Larkin-Imry-Ma state contaminated by the 
network of the topological defects  pinned  by aerogel.}

\begin{document}

\maketitle

\section{Introduction}

Behavior of systems with continuous symmetry in the presence of random
anisotropy disorder is the subject of numerous theoretical and
experimental investigations.
This is because of the surprizing observation
made by Larkin  \cite{Larkin} and Imry and Ma 
\cite{ImryMa} that even a
weak disorder may destroy 
 the long-range  translational or orientational order.
 Recent example is provided by the nematic liquid crystals in
random porous medium, in which the order parameter -- the unit
vector $\hat{\bf n}$ -- interacts with the quenched random
anisotropy disorder (see e.g. Ref.
\cite{Feldman} and references therein). 
Though the problem of violation of the long-range order by
quenched disorder is more than 30 years old, still there is no
complete understanding  (see e.g. Refs.
\cite{Feldman,Nattermann,Wehr,Itakura} 
and references therein),
especially concerning the role of topological defects.

In the anisotropic phase A of superfluid $^3$He, the Larkin-Imry-Ma effect is even more interesting. In this superfluid the order
parameter contains two  Goldstone vector fields: (1) the unit vector
$\hat{\bf l}$ characterizes the spontaneous anisotropy of the orbital
and superfluid properties of the system; and (2) the unit vector $\hat{\bf d}$
characterizes the spontaneous anisotropy of the spin (magnetic) degrees
of freedom. In aerogel, the quenched random anisotropy disorder of the
silicon strands  interacts with the orbital vector $\hat{\bf l}$, which 
thus must experience the Larkin-Imry-Ma effect. As for the
vector $\hat{\bf d}$ of the spontaneous anisotropy of spins it is assumed 
that $\hat{\bf d}$ does not interact directly with  the quenched disorder, at least in the arrangement  when the aerogel strands are
covered by layers of $^4$He atoms preventing  the formation of  solid
layers of  $^3$He with large Curie-Weiss magnetization. There is
a   tiny spin-orbit coupling between vectors $\hat{\bf d}$ and $\hat{\bf l}$ due
to which the $\hat{\bf l}$-vector may transfer
the disorder to the $\hat{\bf d}$-field.  Superfluid $^3$He-A experiences many
different types of topological defects (see e.g. \cite{Book}), which may be pinned by the disorder. 

On recent experiments on the superfluid
$^3$He-A in aerogel see Refs. 
\cite{Nazaretski,experiment,BarkerNegativeShift,Dmitriev1,Dmitriev2,DmitrievNew}
and references therein. In particular, Refs. \cite{experiment,DmitrievNew} describe the transverse
NMR experiments, in which the dependence of the frequency shift on the
tipping angle $\beta$  of the precessing magnetization has been measured;
and Ref. \cite{DmitrievNew} also reports the observation of the longitudinal NMR. 
Here we discuss these experiments in terms of the Larkin-Imry-Ma
disordered state \cite{Larkin,ImryMa} extended for the description of 
the superfluid $^3$He-A in aerogel \cite{Volovik}. 
In Sec. 2  the general equations for NMR in $^3$He-A are written. In Sec. 3 these equations are applied to the states with disordered $\hat{\bf d}$ and $\hat{\bf l}$ fields.
In Sec. 4 and Sec. 5 the models for the two observed states are suggested in terms of the averaged distributions of $\hat{\bf d}$ and $\hat{\bf l}$ fields consistent with observations,. Finally in Sec. 6 these states are interpreted in terms of different types of disorder. 

\section{Larmor precession of $^3$He-A}

In a typical  experimental arrangement the spin-orbit (dipole-dipole) 
energy is  smaller  than  Zeeman energy and thus may be considered as a
perturbation. In zero-order approximation when the  dipole energy  and 
dissipation are neglected, the spin  freely precesses with the Larmor
frequency
$\omega_L=\gamma H$, where $\gamma$ is the gyromagnetic ratio of $^3$He
nuclei. In terms of the Euler angles the precession of magnetization is
given by
\begin{equation}
{\bf S}(t)=S ~R_z(-\omega_L t+\alpha)
 R_y(\beta) \hat {\bf z} ~~.
\label{S}
 \end{equation}
Here $S=\chi H$ is the amplitude of spin induced by magnetic field; the
axis
$\hat {\bf z}$ is along the magnetic field
${\bf H}$; matrix
$R_y(\beta)$ describes rotation about transverse  axis $y$ by angle
$\beta$, which  is the tipping angle of the precessing magnetization; 
$R_z$ describes rotation about  $z$; $\alpha$ is the phase of the
precessing magnetization. According to the Larmor theorem
\cite{BunkovVolovik}, in the precessing frame the vector  $\hat{\bf d}$ 
is in turn precessing about 
${\bf S}$. Because of the interaction between the spin ${\bf S}$ and
the order parameter vector $\hat{\bf d}$, the precession of $\hat{\bf d}$
occurs  in the plane perpendicular to ${\bf S}$,  and it is characterized
by another phase
$\Phi_d$. In the laboratory frame the precession of $\hat{\bf d}$ is given by
\begin{equation}
\hat {\bf d}(t)=R_z(-\omega_L t+\alpha)
 R_y(\beta) R_z(\omega_L t-\alpha+\Phi_d)\hat {\bf x} ~, 
\label{d}
 \end{equation}
while the orbital vector  $\hat{\bf l}$ is time-independent in this approximation:
\begin{equation}
\hat{\bf l}=\hat{\bf z} \cos\lambda
+ 
\sin\lambda (\hat{\bf x}\cos\Phi_l+\hat{\bf y}\sin\Phi_l) ~.
\label{l}
 \end{equation}
This is the general state of the pure Larmor precession of $^3$He-A, and
it contains 5 Goldsone parameters: 2 angles $\alpha$ and $\beta$ of the
magnetization in the precessing frame; angle
$\Phi_d$ which characterizes the precession of vector $\hat{\bf d}$; and
two angles 
$\lambda$  and $\Phi_l$ of the orbital vector $\hat{\bf l}$.

The degeneracy is lifted by spin-orbit (dipole) interaction 
\cite{VollhardtWolfle}
\begin{equation}
F_D=-\frac{\chi\Omega^2_A}
{2\gamma^2}(\hat{\bf l}\cdot
\hat{\bf d})^2~, 
\label{DipoleInter}
 \end{equation}
  where  $\Omega_A$ is the
so-called Leggett frequency. In the bulk homogeneous $^3$He-A,   the
Leggett frequency coincides with the frequency of the longitudinal NMR,
$\omega_\parallel=
\Omega_A$. In typical experiments one has $\Omega_A\ll
\omega_L$, which allows us to use   the spin-orbit interaction averaged over the fast Larmor
precession:
\cite{Gongadze,BunkovVolovik}
 \begin{eqnarray}
\bar F_D =\frac{\chi\Omega^2_A}
{2\gamma^2} U~~,
\label{SO1} 
\\
U=-{1\over 2}\sin^2\beta + 
{1\over 4}(1+\cos\beta)^2\sin^2\Phi\sin^2\lambda\nonumber\\
-( {7\over
8}\cos^2\beta +{1\over 4}\cos\beta  -{1\over 8})\sin^2\lambda
~~,
\label{SO} 
 \end{eqnarray}
where $\Phi=\Phi_d-\Phi_l$.
    
The dipole interaction generates the frequency shift of the transverse 
NMR from the Larmor frequency:
\begin{equation}
 \omega_\perp- \omega_L=-\frac{\partial \bar F_D}{\partial (S\cos\beta)} 
=- 
\frac{\Omega^2_A}{2\omega_L} \frac{\partial U}{\partial  \cos\beta}~. ~~
\label{FShift} 
 \end{equation}

In the bulk $^3$He-A, the minimum of the dipole interaction requires that
$\Phi_d=\Phi_l$, and $\sin^2\lambda=1$, i.e. the equilibrium
position of $\hat{\bf l}$ is in the plane perpendicular to ${\bf H}$.
However, for the $^3$He-A confined in aerogel, the interaction with the quenched disorder may essentially modify this spatially homogeneous state
by destroying the long-range orientational order due to the Larkin-Imry-Ma 
effect
\cite{Volovik}.

\section{Two states of $^3$He-A in aerogel}

Experiments reported in Ref. \cite{DmitrievNew} demonstrate two different types of 
magnetic behavior of the A-like phase in aerogel, denoted as (f+c)-state and
(c)-state correspondingly. The (f+c)-state contains two overlapping lines (f)  and
(c) in the transverse NMR spectrum ({\it far} from and {\it close} to the Larmor frequency correspondingly). The frequency shift  of the transverse
NMR is about 4 times bigger for the (f)-line compared to the (c)-line. 
The behavior under applied gradient of magnetic field suggests that  the 
(f+c)-state  consists of two magnetic states concentrated in different parts of
the cell. The (c)-state contains  only a single (c)-line in the spectrum, and it is
obtained after application of the 180 degree pulse while cooling through
$T_c$. The pure (f)-state, i.e.
the state with a single (f)-line,  has not been observed. 

The (c) and (f+c) states have different of  dependence of the frequency 
shift $\omega_\perp- \omega_L$ on the tilting angle $\beta$ in the pulsed NMR experiments:  $\omega_\perp-
\omega_L \propto  \cos\beta$ in the pure (c)-state; and
  $\omega_\perp- \omega_L \propto (1+\cos\beta)$ in the (f+c)-state.  The
latter behavior probably characterizes  the property of the (f)-line which has
the bigger shift and is dominating in the spectrum of the (f+c)-state. 
The $(1+\cos\beta)$-law has been also observed in Ref.
\cite{experiment}.

The experiments with longitudinal NMR were also reported  in Ref.
\cite{DmitrievNew}. The  longitudinal resonance in the (f+c)-state has been
observed, however no traces of the longitudinal  resonance have been seen in the
(c)-state. 

Let us discuss this behavior in terms of the disordered states emerging  in
$^3$He-A due to the orientational disorder.

In the extreme case of weak disorder,  the characteristic Imry-Ma length
$L$ of the disordered $\hat{\bf l}$-texture is much bigger than the
characteristic length scale  $\xi_D$ of dipole interaction, $L\gg \xi_D$.
In this case  the equilibrium values of
$\Phi$ and $\lambda$ are dictated by the spin-orbit interaction, $\Phi=\Phi_d-\Phi_l=0$ and  $\sin^2\lambda=1$; and the Eq.(\ref{FShift}) gives
\begin{equation}
\omega_\perp- \omega_L=   \frac{\Omega^2_A}{8\omega_L} (1+3  \cos\beta)~.
\label{case1} 
 \end{equation}
This dependence fits neither the (f)-state $(1+\cos\beta)$ behavior nor  the
$\cos\beta$ law in the (c)-state, which indicates that the disorder is not weak compared to the spin-orbit energy.
 
In the extreme case of strong disorder, when $L\ll \xi_D$, both $\Phi$
   and  $\hat{\bf l}$ are random:
\begin{eqnarray}
 \left<\sin^2\Phi\right>-\frac{1}{2} =0~,
\label{randomPhi} 
\\
\left<\sin^2\lambda\right>
-\frac{2}{3}=0~.
\label{randomLambda} 
 \end{eqnarray} 
 In this case
it follows from Eq.(\ref{FShift}) that the frequency shift is absent:
\begin{equation}
\omega_\perp= \omega_L~.
\label{case2} 
 \end{equation} 
The Eq.(\ref{randomPhi}) means that $\Phi_l$ and $\Phi_d$ are dipole
unclocked, i.e. they are not locked by the spin-orbit
dipole-dipole interaction, which is natural in case of small Imry-Ma
length, $L\ll
\xi_D$.  In principle, there can be three different dipole-unlocked cases: (i) 
when  both
$\Phi_l$ and
$\Phi_d$ are random and independent; (ii)  when $\Phi_l$ is random while
$\Phi_d$ is fixed; (iii)  when $\Phi_d$ is random while $\Phi_l$ is fixed.

The strong disorder limit is consistent with the observation that the frequency shift of the (c)-line is much smaller
than the frequency shift in $^3$He-B in aerogel. The observed non-zero value can be explained in terms of the small first order corrections to the strong disorder limit.  Let us 
introduce the parameters
\begin{equation}
a=\frac{1}{2} -\left<\sin^2\Phi\right>~~,~~b=\left<\sin^2\lambda\right>
-\frac{2}{3}~,
\label{random2} 
 \end{equation}
which describe the deviation from the strong disorder limit.
These parameters are zero in the limit of strong disorder $L^2/\xi_D^2\rightarrow 0$, and one may expect that in the pure Larkin-Imry-Ma state they are proportional to the small parameter $L^2/\xi_D^2\ll 1$.
The behavior of these two parameters can be essentially different  in different
realizations  of the disordered state, since the vector $\hat{\bf l}$ entering the parameter $a$  interacts
with the quenched orientational disorder directly, while $\Phi_d$  only interacts with
$\Phi_l$ via the spin-orbit coupling. That is why we shall try to interpret the two observed magnetic states,
the  (c)-state and the (f)-state, in terms of different realizations of the textural disorder described by different phenomenological relations between
parameters $a$ and $b$ in these two states.

\section{Interpretation of (c)-state}
\label{cstate}

The  observed $\cos\beta$-dependence of the transverse NMR  frequency shift in the
(c)-state \cite{DmitrievNew} can be reproduced if  we assume that in the (c)-state
the parameters $a_c$ and $b_c$ satisfy the following relation: $a_c\ll b_c$. Then in the main approximation,
\begin{equation}
a_c=0~~,~~b_c\neq 0~,
\label{Stateb} 
 \end{equation}
the effective potential $U$ in Eq.(\ref{SO}) is
\begin{equation}
U_c=- 
\frac{3}{4}b_c \cos^2\beta+\frac{1}{4}\left(b_c +\frac{1}{3}\right)~. 
\label{PotentialRandom2} 
 \end{equation}
If the parameter $b_c$ does not depend on $\beta$, the variation of  $U_c$
with respect to $\cos\beta$  gives  the required $\cos\beta$-dependence of the 
frequency shift of transverse NMR in the (c)-state:
\begin{equation}
 \omega_{c\perp}- \omega_L= 
\frac{3\Omega^2_A}{4\omega_L }b_c\cos\beta~. 
\label{TransverseRandom2} 
 \end{equation}

 Let us estimate  the parameter $b_c$ using the following consideration. 
The dipole energy which depends on $\lambda$ violates the complete 
randomness of the Larkin-Imry-Ma state, and thus perturbs the average value of
$\sin^2\lambda$. The deviation of $b_c$ from zero is given by:
\begin{equation}
b_c \sim \frac{L^2}{\xi_D^2}\left(\cos^2\beta
-\frac{1}{3}\right)~.
\label{CorrectionRandom2} 
 \end{equation}
In this model the  potential and frequency shift become
\begin{equation}
U_c\sim - \frac{L^2}{\xi_D^2}\left(\cos^2\beta
-\frac{1}{3}\right)^2~, 
\label{PotentialRandom2Final} 
 \end{equation} 
\begin{equation}
  \omega_{c\perp}- \omega_L\sim  
\frac{\Omega^2_A}{\omega_L} \frac{L^2}{\xi_D^2}\cos\beta
\left(\cos^2\beta
-\frac{1}{3}\right)~. 
\label{CubicTransverseRandom2} 
 \end{equation}
Such $\beta$-dependence of the transverse NMR is also antisymmetric with respect to
the transformation
$\beta\rightarrow \pi-\beta$ as in the model with the $\beta$-independent parameter
$b_c$ in Eq.(\ref{TransverseRandom2}); however, as distinct from that model it is
inconsistent with the experiment (see Fig.~6 in Ref.
\cite{DmitrievNew}). Certainly the theory must be refined to estimate the first
order corrections to the zero values of the parameters
$a_c$ and $b_c$.

The frequency of the longitudinal NMR in the (c)-state  is zero in the
local approximation. The correction due to the deviation of $\Phi$
from the random behavior, i.e. due to the non-zero value  of  the parameter
$a_c$ in the (c)-state, is:
\begin{equation}
\omega_{c\parallel}^2 =   
\frac{2}{3}\left(1-2\left<\sin^2\Phi\right>\right)\Omega^2_A =
\frac{4a_c}{3} \Omega^2_A~.
\label{LongitudinalNMRStrong} 
 \end{equation}
In the simplest Imry-Ma model  $a_c \sim  L^2/\xi_D^2 \ll 1$,  and thus the
frequency of longitudinal NMR is small as compared with the  frequency of the
longitudinal resonance in (f+c)-state, discussed in the next section. This is
consistent with non-observation of the longitudinal NMR in the (c)-state: under
conditions of this experiment  the longitudinal resonance cannot be seen if its
frequency is  much smaller than the frequency of the  longitudinal resonance
observed in the (f+c)-state  \cite{DmitrievNew}.   

\section{Interpretation of (f)-state}
\label{fstate}

The observed $(1+\cos\beta)$-dependence of the transverse NMR frequency  shift of
the (f)-line dominating in the (f+c)-state \cite{DmitrievNew,experiment} is reproduced if  we
assume that for the (f)-line one has
$a_f\gg b_f$. In this case, in the main approximation the (f)-state may be 
characterized by
\begin{equation}
a_f \neq 0~~,~~b_f=0~,
\label{random3} 
 \end{equation}
and one obtains:
\begin{equation}
\omega_{f\perp}- \omega_L= \frac{a_f}{6} \frac{\Omega^2_A}{\omega_L}  (1+ 
\cos\beta)~.
\label{aStateTransverse} 
 \end{equation}
Let us compare the frequency shift  of the (f)-line with that of the  (c)-line  in
Eq. (\ref{TransverseRandom2})  at $\beta=0$:
\begin{equation}
\frac{\omega_{f\perp}- \omega_L}{\omega_{c\perp}- \omega_L} =\frac{4a_f}{9b_c} ~.
\label{f/c} 
 \end{equation}
According to experiments \cite{DmitrievNew}  this ratio is about 4, and thus one obtains  the
estimate:  $b_c \sim 0.1 a_f$. This supports the strong disorder limit, $b_c\ll 1$, for the (c)-state. If the statistic properties of the $\hat{\bf
l}$-texture in  the (f)-state are similar to that in the (c)-state, then one has
$b_f\ll a_f$ as suggested in Eq.(\ref{random3}).

The frequency of the longitudinal NMR in such a state  is
\begin{equation}
\omega_{f\parallel}^2 =   \frac{4a_f}{3}\Omega^2_A ~,
\label{aStateLongitudinal} 
 \end{equation}
which gives the relation between the transverse and  longitudinal NMR
frequencies
\begin{equation}
\omega_{f\perp}- \omega_L= \frac{1}{8} \frac{\omega_{f\parallel}^2}{\omega_L}  
(1+ 
\cos\beta)~. 
\label{UniversalRelation} 
 \end{equation} 
 This relation  is also valid for the Fomin's robust phase  \cite{Fomin} (see 
\cite{ResultForRobust}). However, the frequency of the  longitudinal NMR measured
in the (f+c)-state  \cite{DmitrievNew} does not satisfy this  relation: the
measured value of $\omega_{f\parallel}$ is about 0.65 of the value which follows
from the Eq.(\ref{UniversalRelation}) if one uses the measured $\omega_{f\perp}-
\omega_L$.   Probably the situation can be improved, if one considers the
interaction between the $f$ and $c$ lines in the (f+c)-state (see Ref.
\cite{DmitrievNew}).

\section{Discussion}

\subsection{Interpretation of A-phase states in aerogel.}

The observed two magnetic states of $^3$He-A  in aerogel \cite{DmitrievNew} can be
interpreted in the following way. The pure (c)-state is the Larkin-Imry-Ma phase
with strong disorder, $L\ll
\xi_D$.  The (f+c)-state can be considered as mixed state with the volume $V_c$  
filled by the Larkin-Imry-Ma phase, while the rest of volume $V_f=V-V_c$ consists
of  the (f)-state. The (f)-state is also random due to the Larkin-Imry-Ma effect, 
but  the spin variable  $\Phi_d$ and the orbital variable $\Phi_l$ are not
completely independent in this state. If $\Phi_d$   partially follows
$\Phi_l$, the difference $\Phi=\Phi_d-\Phi_l$ is not random  and the parameter 
$a_f$  in the (f)-state is  not very small, being equal to $1/2$ in the extreme
dipole-locked case. Thus, for the (f+c)-state one may assume that $a_f\gg b_f,
b_c,a_c$.  As a result the (f)-line has essentially larger frequency shift  of
transverse NMR and essentially larger longitudinal frequency compared to the
(c)-line. 

Both results are consistent with the experiment:  from the transverse NMR   it
follows that $b_c \sim 0.1 a_f$ (see Eq.(\ref{f/c})); and from the lack of the
observation of longitudinal NMR  in the (c)-state it follows that $a_c \ll a_f$.
This confirms the assumption of the strong disorder in the (c)-state, in which the
smallness of the parameters
$b_c$ and $a_c$ is the result of the randomness of the $\hat{\bf l}$-texture  on
the length scale $L\ll\xi_D$.
The $\cos\beta$-dependence of $\omega_{c\perp}- \omega_L$ in
Eq.(\ref{TransverseRandom2}) and 
$(1+\cos\beta)$-dependence of $\omega_{f\perp}- \omega_L$ in
Eq.(\ref{aStateTransverse}) are also consistent with the experiment. The open problem is how to estimate theoretically the phenomenological parameters
$a_f,b_f, b_c,a_c$ and find its possible dependence on $\beta$. 

The `universal' relation (\ref{UniversalRelation}) 
between  the longitudinal and
transverse NMR frequencies is not satisfied in the experiment,  but we cannot
expect the exact relation in such a crude model, in which  the interaction between
the $f$ and $c$ lines in the (f+c)-state is ignored (see \cite{DmitrievNew}).
Moreover, we use the local approximation,  i.e. we do not take into account the
fine structure of the NMR line which may contain the satellite peaks due to the
bound states  of spin waves in the texture of
$\hat{\bf l}$ and $\hat{\bf d}$ vectors. The tendency is however correct:  the
smaller is the frequency shift  of transverse NMR the smaller is the frequency of
longitudinal NMR.


\subsection{Global anisotropy and negative frequency shift}

For further consideration one must take into account that  in some aerogel
samples the large negative frequency shift has been observed for the
A-phase 
\cite{BarkerNegativeShift,Dmitriev1,Dmitriev2,BunkovPrivate}. 
The reason of the negative shift is the deformation
of the aerogel sample which leads to  the global 
orientation of  the orbital vector $\hat{\bf l}$ in the large region of
the aerogel \cite{BunkovPrivate}. The effect of regular uniaxial
anisotropy in aerogel has been considered in Refs. \cite{Pollanen,Aoyama}.
It is important that even a rather small deformation of aerogel may
kill the subtle collective Larkin-Imry-Ma effect and lead to the uniform
orientation of the
$\hat{\bf l}$-vector.  Using the estimation of the Imry-Ma length in Ref.
\cite{Volovik}, one can find that the critical stretching of the aerogel
required to kill the Larkin-Imry-Ma effect is proportional to
$(R/\xi_0)^3$. Here
$R$ is the radius of the silica strands and $\xi_0$ is the superfluid
coherence length.

 From
Eqs.(\ref{SO}) and (\ref{FShift}) it follows that the maximum possible
negative frequency shift could occur if in some region  the global
orientation of 
$\hat{\bf l}$ induced by deformation of the aerogel is along the magnetic
field (i.e. $\lambda=0)$:
\begin{equation}
\omega_\perp- \omega_L= - \frac{\Omega^2_A}{2\omega_L} ~.
\label{negative} 
 \end{equation}
Such longitudinal orientation of $\hat{\bf l}$ is possible because the
regular anisotropy caused by the deformation of aerogel is bigger than the
random anisotropy, which in turn in the strong disorder limit  is bigger
than the dipole energy preferring the transverse orientation of
$\hat{\bf l}$. 

Comparing
the measured magnitude of the negative shift  (which cannot be bigger
than the maximum possible in Eq.(\ref {negative})) with  the measured
positive shift of the (f)-line in the (f+c)-state 
\cite{Dmitriev,DmitrievNew}  one obtains that the parameter $a_f$ in
Eq.(\ref{aStateTransverse}) must be smaller than $0.25$. This is also
confirmed by the results of   the longitudinal NMR experiments
\cite{DmitrievNew}, which show  that the frequency  of the longitudinal
NMR in the (f+c)-state of $^3$He-A is much smaller than  the frequency 
of the longitudinal NMR in $^3$He-B. The latter  is only possible if
$a_f\ll 1$ in Eq.(\ref{aStateLongitudinal}), i.e. the (f)-state is also in the
regime of strong disorder. Thus there is only the partial dipole locking between
the spin variable  $\Phi_d$ and the orbital variable $\Phi_l$ in the (f)-state.

\subsection{Possible role of topological defects.}

It is not very clear what is the origin of the (f)-state. The partial
dipole locking is possible if  the characteristic size of the $\hat{\bf
l}$ texture in the (f)-state is on the order of or somewhat smaller   than
$\xi_D$. 

Alternatively, the line (f) could come from the topological defects of
the A-phase (vortices, solitons, vortex sheets,  etc., see Ref.
\cite{Book}). The defects could appear during cooling down the sample
from the normal (non-superfluid) state and are annealed by application of
the 180 degree pulse during this process. Appearance of a large amount
of pinned topological defects in $^3$He-B in aerogel has been suggested
in Ref.
\cite{Collin}. The reason why the topological defects may effect the NMR
spectrum in $^3$He-A is the following. In  the case of the  strong
disorder limit the texture is random, and the frequency shift is zero, if
one neglects the $ L/\xi_D$ corrections in the main approximation.  The
topological defect introduces some kind of order: some correlations are
nonzero because of the conserved topological charge of the defect. That
is why the frequency shift will be nonzero. It will be small, but still
bigger than due to the corrections of order  
$(L/\xi_D)^2$.

If this interpretation is correct, there are two different realizations
of the disordered state in the system with quenched orientational
disorder: the network of the pinned topological defects and the pure
Larkin-Imry-Ma state. Typically one has the interplay between these two
realizations, but the defects can be erased by the proper annealing
leaving  the pure Larkin-Imry-Ma state.

\subsection{Superfluid properties of A-phase in aerogel}

The interesting problem concerns the superfluid density $\rho_s$  in the
states with the orientational  disorder in the vector $\hat{\bf l}$. In Ref. 
\cite{Volovik}  it was suggested that
$\rho_s=0$ in such a state. Whether the superfluid density is zero or not
depends on the rigidity of the $\hat{\bf l}$-vector. If the  $\hat{\bf
l}$-texture is flexible, then due to the Mermin-Ho relation between the
texture and the superfluid velocity, the texture is able to respond to
the applied superflow by screening the supercurrent. As a result
the superfluid density in the flexible texture could be zero.   The
experiments on $^3$He-A in aerogel demonstrated that  $\rho_s\neq 0$ (see
e.g.
\cite{Nazaretski} and references therein). 

However, most probably  these
experiments have been done in the (f+c)-state. If our interpretation of this
state in terms of the topological defects is correct, the non-zero value of
superfluid density could be explained in terms of the pinning of the  defects
which leads to the effective rigidity  of the $\hat{\bf l}$-texture  in the
(f+c)-state.  
Whether the superfluid density is finite in the pure Larkin-Imry-Ma
state, identified here as the (c)-state, remains an open experimental and theoretical
question.  The theoretical discussion of the rigidity or quasi-rigidity in such a
state can be found in Refs.
\cite{EfetovLarkin,Itakura}. In any case, one may expect that the observed two
states of $^3$He-A in aerogel, (c) and (f+c), have different superfluid
properties. 

Recent Lancaster experiments with vibrating aerogel sample indicate
that the sufficiently large superflow produces the state with the regular (non-random) orientation of $\hat{\bf l}$ in aerogel, and in the
oriented state the superfluid density is bigger \cite{Fisher}.
This suggests that the orientational disorder does lead to at least partial suppression of the superfluid density.

\subsection{Conclusion.}

In conclusion, the NMR experiments on the A-like superfluid state   in
the aerogel indicate two types of behavior. Both of them can be
interpreted in terms of the random  texture of the orbital vector
$\hat{\bf l}$ of the $^3$He-A order parameter. This supports the idea
that  the superfluid  $^3$He-A in aerogel exhibits the  Larkin-Imry-Ma
effect: destruction of the long-range orientational order by random
anisotropy produced by the randomly oriented silicon strands of the aerogel. The
extended numerical simulations are needed to clarify the role of the topological
defects in the Larkin-Imry-Ma state, and to calculate the dependence of the NMR
line-shape and superfluid density on concentration and pinning of the
topological defects.

I thank V.V. Dmitriev and D.E. Khmelnitskii for illuminating discussions,
 V.V. Dmitriev, Yu.M. Bunkov and S.N. Fisher  for presenting the experimental results before
publication, and I.A. Fomin who attracted my attention to the relation
between frequencies of  the longitudinal and transverse NMR in some
states. This work was supported in part by the Russian Foundation for
Fundamental Research and the ESF Program COSLAB.

\end{document}